\documentclass[12pt,preprint]{aastex}

\newcommand{\h}{H1426+428 }

\newcommand{\hh}{H1426+428}

\newcommand{\C}{\v Cerenkov\ }

\newcommand{\on}{\textsc{on }}
\newcommand{\onn}{\textsc{on}}
\newcommand{\off}{\textsc{off }}
\newcommand{\offf}{\textsc{off}}
\newcommand{\onoff}{\textsc{on/off }}

\newcommand{\tracking}{\textsc{tracking }}
\newcommand{\trackingg}{\textsc{tracking}}
\newcommand{\pairs}{\textsc{pairs }}
\newcommand{\pairss}{\textsc{pairs}}

\newcommand{\bpsp}{\emph{Beppo}SAX }

\newcommand{\gsp}{$\gamma$ }

\newcommand{\gr}{$\gamma$-ray }
\newcommand{\grr}{$\gamma$-ray}
\newcommand{\grs}{$\gamma$-rays }
\newcommand{\grss}{$\gamma$-rays}
\newcommand{\dg}{\ensuremath{^\circ} }
\newcommand{\dgg}{\ensuremath{^\circ}}

\newcommand{\al}{\emph{alpha }}

\newcommand{\ep}{E$_{peak}$ }
\newcommand{\epp}{E$_{peak}$}

\newcommand{\bls}{BL Lacs }

\shorttitle{Detection of H1426+428 at TeV \gr energies}
\shortauthors{Horan D., et al.}

\begin{document}

\title{Detection of the BL Lac Object H1426+428 at TeV Gamma Ray Energies}

\author{D. Horan,\altaffilmark{1,2}
H. M. Badran,\altaffilmark{1,7}
I. H. Bond,\altaffilmark{3}
S. M. Bradbury,\altaffilmark{3}
J. H. Buckley,\altaffilmark{4}
M. J. Carson,\altaffilmark{2}
D. A. Carter-Lewis,\altaffilmark{5}
M. Catanese,\altaffilmark{1}
W. Cui,\altaffilmark{7}
S. Dunlea,\altaffilmark{2}
D. Das,\altaffilmark{15}
I. de la Calle Perez,\altaffilmark{3}
M. D'Vali,\altaffilmark{3}
D. J. Fegan,\altaffilmark{2}
S. J. Fegan,\altaffilmark{1,8}
J. P. Finley,\altaffilmark{6}
J. A. Gaidos,\altaffilmark{6}
K. Gibbs,\altaffilmark{1}
G. H. Gillanders,\altaffilmark{9}
T. A. Hall,\altaffilmark{5,17}
A. M. Hillas,\altaffilmark{3}
J. Holder,\altaffilmark{3}
M. Jordan,\altaffilmark{4}
M. Kertzman,\altaffilmark{16}
D. Kieda,\altaffilmark{11}
J. Kildea,\altaffilmark{2}
J. Knapp,\altaffilmark{3}
K. Kosack,\altaffilmark{4}
F. Krennrich,\altaffilmark{5}
M. J. Lang,\altaffilmark{9}
S. LeBohec,\altaffilmark{5}
R. Lessard,\altaffilmark{6}
J. Lloyd-Evans,\altaffilmark{3}
B. McKernan,\altaffilmark{2}
P. Moriarty,\altaffilmark{13}
D. Muller,\altaffilmark{10}
R. Ong,\altaffilmark{12}
R. Pallassini,\altaffilmark{3}
D. Petry,\altaffilmark{5}
J. Quinn,\altaffilmark{2}
N. W. Reay,\altaffilmark{15} 
P. T. Reynolds,\altaffilmark{14}
H. J. Rose,\altaffilmark{3}
G. H. Sembroski,\altaffilmark{6}
R. Sidwell,\altaffilmark{15}
N. Stanton,\altaffilmark{15}
S. P. Swordy,\altaffilmark{10}
V. V. Vassiliev,\altaffilmark{11}
S. P. Wakely,\altaffilmark{10}
T. C. Weekes\altaffilmark{1}}

\email{dhoran@cfa.harvard.edu} 

\altaffiltext{1}{Fred Lawrence Whipple Observatory, Harvard-Smithsonian CfA, P.O. Box 97, Amado, AZ 85645} 
\altaffiltext{2}{Experimental Physics Department, National University of Ireland, Belfield, Dublin 4, Ireland}
\altaffiltext{3}{Department of Physics, University of Leeds, Leeds, LS2 9JT, Yorkshire, England, UK}
\altaffiltext{4}{Department of Physics, Washington University, St. Louis, MO 63130}
\altaffiltext{5}{Department of Physics and Astronomy, Iowa State University, Ames, IA 50011}
\altaffiltext{6}{Department of Physics, Purdue University, West Lafayette, IN 47907}
\altaffiltext{7}{Physics Department, Tanta University, Tanta, Egypt}
\altaffiltext{8}{Department of Physics, University of Arizona, Tucson, AZ 85721}
\altaffiltext{9}{Physics Department, National University of Ireland, Galway, Ireland}
\altaffiltext{10}{Enrico Fermi Institute, University of Chicago, Chicago, IL 60637}
\altaffiltext{11}{High Energy Astrophysics Institute, University of Utah, Salt Lake City, UT 84112}
\altaffiltext{12}{Department of Physics, University of California, Los Angeles, CA 90095}
\altaffiltext{13}{School of Science, Galway-Mayo Institute of Technology, Galway, Ireland}
\altaffiltext{14}{Department of Physics, Cork Institute of Technology, Cork, Ireland}
\altaffiltext{15}{Department of Physics, Kansas State University, Manhattan, KS 66506}
\altaffiltext{16}{Department of Physics and Astronomy, DePauw University, Greencastle, IN 46135}
\altaffiltext{17}{Physics \& Astronomy Department, University of Arkansas at Little Rock, Little Rock, AR 72204}

\begin{abstract}

A very high energy \gr signal has been detected at the 5.5$\sigma$
level from \hh, an x-ray selected BL Lacertae object at a redshift of
0.129. The object was monitored from 1995 - 1998 with the Whipple 10m
imaging atmospheric \C telescope as part of a general blazar survey;
the results of these observations, although not statistically
significant, were consistently positive. X-ray observations of \h
during 1999 with the \bpsp instrument revealed that the peak of its
synchrotron spectrum occurs at $>$ 100 keV, leading to the prediction
of observable TeV emission from this object. \h was monitored
extensively at the Whipple Observatory during the 1999, 2000, and 2001
observing seasons. The strongest TeV signals were detected in 2000 and
2001. During 2001, an integral flux of 2.04 $\pm$ 0.35 x 10$^{-11}$
cm$^{-2}$ s$^{-1}$ above 280 GeV was recorded from \hh. The detection
of \h supports the idea that, as also seen in Markarian 501 and
1ES2344+514, BL Lacertae objects with extremely high synchrotron peak
frequencies produce \grs in the TeV range.

\end{abstract}


\section{Introduction}

Blazars are the main class of Active Galactic Nuclei (AGN) detected
above 100 MeV by the EGRET experiment on the Compton Gamma Ray
Observatory (CGRO) and by ground-based \gr observatories
\citep{Mukherjee:97,Weekes:01a}. They comprise a subclass of AGN and
are characterized by a highly variable non-thermal continuum, strong
variable optical polarization, the lack of a UV-excess (or ``blue
bump'') and a core-dominated radio morphology. BL Lacertae (BL Lac)
objects are a subclass of blazars that are notable for their lack of
prominent emission lines. The broad-band double-humped Spectral Energy
Distributions (SEDs) of BL Lacs identified in x-ray surveys differ
significantly from the SEDs of those identified in radio surveys. This
led to the sub-classification of BL Lacs into High-frequency-peaked BL
Lacs (HBLs) and Low-frequency-peaked BL Lacs (LBLs) based on the ratio
of their x-ray to radio flux densities \citep{Padovani:95}. The first
``hump'', generally assumed to be the peak of the synchrotron emission
(in a $\nu F_{\nu}$ representation), is in the IR-optical for LBLs and
in the EUV-soft x-ray band for HBLs. BL Lac objects make up a
significant fraction of the 70 blazars in the 3rd EGRET Catalog
\citep{Hartman:99} and most of them are classified as LBLs. It has been 
shown that there is not a sharp division between these two classes of
objects (see, e.g., \citealp{Fossati:98}; \citealp{Ghisellini:99}).

Recent studies with ground-based \gr telescopes have produced evidence
for TeV \gr emission from seven BL Lac objects, five of which are
classified as HBL and two as LBL \citep{Weekes:01a}.  The most
prominent of these are Markarian 421 (Mrk 421), which has been
detected by five ground based imaging atmospheric \C \gr
observatories, and Markarian 501 (Mrk 501), which has been detected by
six such observatories. The emission from both of these objects can be
explained by Compton-synchrotron models although detailed modeling is
still fraught with many uncertainties. Both objects are characterized
by rapid variability on time-scales from hours to months. In the TeV
energy range, the energy spectrum of Mrk 501 and, more recently, that
of Mrk 421, have been shown to exhibit absorption-like features
\citep{Krennrich:99,Aharonian:99,Krennrich:01}.  The temporal and
spectral properties of the other TeV BL Lacs are less well defined.

Since 1992, the Whipple Gamma Ray collaboration, using the 10m imaging
atmospheric \C telescope on Mt. Hopkins, has been searching for TeV
\gr emission from AGN. Initially the search was concentrated on
blazars detected by EGRET at any redshift; these observations led to
the detection of Mrk 421 \citep{Punch:92} and upper limits on some 30
other blazars \citep{Kerrick:95}. More recently, the search has
concentrated on nearby BL Lacs leading to the detection of Mrk 501
\citep{Quinn:96} and 1ES 2344+514 \citep{Catanese:98}. Between 1995
and 1998, the survey included 24 objects (17 HBLs, 7 LBLs) ranging in
redshift from 0.046 to 0.44; the results for these objects will be
published shortly (D. Horan et al., in preparation). Although none of
these observations resulted in a detection, the observations of \h
yielded the highest consistently positive statistical significances.

In this paper, evidence is presented for the detection of very high
energy (VHE) \grs from \hh. The \h observations are divided into two
categories - those taken as part of the general blazar survey between
1995 and 1998, and the subsequent concentrated observations carried
out between 1999 and 2001. A preliminary energy spectrum is derived
and the implications of the detection of H1426+428 at these energies
are discussed.

\section{H1426+428 at Other Wavelengths}

H1426+428 was discovered in the 2-6 keV band by HEAO-1
\citep{Wood:84} and was classified as a BL Lac object in 1989
\citep{Remillard:89}. It has a cosmological redshift of 0.129 and is
located in the constellation of B\"{o}otes ($\alpha_{J2000}$ = 14$^h$
28$^{m}$ 32$^{s}$.7, $\delta_{J2000}$ = +42$^{\circ}$
40$^{\prime}$ 20$^{\prime\prime}$). It is an optically faint object
(m$_V$ = 16.9) and is believed to be at the center of an elliptical
galaxy. \h is bright in the x-ray band with a 2-6 keV luminosity of
$\sim$ 10$^{44}$ erg s$^{-1}$ which is typical of the \bls found with
HEAO-1. Both the flux of H1426+428 in the 2-10 keV band and its
spectral index above 2 keV have been found to change over time
\citep{Costamante:01}.

Three recent BL Lac surveys, DXRBS \citep{Perlman:98}, RGB
\citep{Laurent:99}, and REX \citep{Caccianiga:99}, have shown that,
despite the conventional subdivision into HBLs and LBLs, BL Lacs
actually form a continuous class with respect to the peak of the
synchrotron emission, which smoothly ranges between IR and soft x-ray
frequencies and up to the 2-10 keV band for sources like Mrk
421. {\it{Beppo}}-SAX observations of Mrk 501 \citep{Pian:98} and 1ES
2344+514 \citep{Giommi:00} revealed that, at least in a flaring state,
the first peak of the SED can reach even higher energies, at or above
100 keV.

In 1998-1999, {\it{Beppo}}-SAX performed an observing campaign with
the aim of finding and studying other sources as ``extreme'' as Mrk
501 is in its flaring state. The candidates for the {\it{Beppo}}-SAX
survey were selected from the Einstein Slew Survey and the RASSBSC
catalogs. These {\it{Beppo}}-SAX observations (Costamante et
al. 2000a, 2000b, 2001) revealed four new ``extreme'' HBLs, selected
because they have high synchrotron peak frequencies and are therefore
possible TeV emitters. These four candidates for TeV emission were:
1ES 0120+340, PKS 0548-322, 1ES 1426+42.8 (i.e. H1426+428) and
H2356-309.

The spectra for three of these objects (1ES 0120+340, PKS 0548-322,
and H2356-309) were well fitted by a convex broken power law with a
break energy, and hence the peak of the synchrotron emission,
occurring at about 1.4 keV for 1ES0120+340, 4.4 keV for PKS0548-322
and 1.8 keV for H2356-309. In the case of H1426+428 however, no
evidence for a spectral break up to 100 keV was found. Instead, its
spectrum was well fitted by a single power-law, with a flat spectral
index of 0.92 up to 100 keV, thus constraining the peak of the
synchrotron emission to lie near or above this value. The best fit of
a pure homogeneous synchrotron self-Compton (SSC) model for H1426+428
\citep{Costamante:00a} predicted detectable \gr emission at TeV
energies. At the time of the {\it {Beppo}}-SAX observation the
observed x-ray flux in the 2-10 keV band was at one of the lowest
levels ever recorded from \hh, indicating that it was not in a flaring
state \citep{Costamante:01}. This implies that, in the event of a
flare, the synchrotron peak could shift to even higher values as was
observed for both Mrk 501 and 1ES 2344+514. It also indicates that
highly relativistic electrons are present which makes H1426+428 the
most promising candidate for TeV emission from this survey.

\section{Observing Technique}

The observations reported in this paper were taken with the 10m
reflector at the Whipple Observatory on Mount Hopkins in southern
Arizona (elevation 2.3 km) using the atmospheric \C imaging
technique. During the course of the observations presented here, many
changes were implemented on the 10m telescope. The imaging camera
\citep{Cawley:90} was upgraded a number of times \citep{Finley:00}, 
the number of photo-multiplier tubes (PMTs) being increased with each
iteration, resulting in the current high resolution, 490 pixel camera
\citep{Finley:01}. The triggering electronics were upgraded to a
Pattern Selection Trigger \citep{Bradbury:99} and light cones were
installed in front of the PMTs to increase their light collection
efficiency. The unprotected, anodized, front-aluminized mirrors on the
10m were recoated during the timespan of the observations. Because of
changes in PMT configuration, triggering, mirror reflectivity, and
light collection efficiency, the energy response of the camera for \gr
detection varied. Since various changes were made at the beginning of
each observing season (typically between June and September), it is
convenient to consider each observing season separately and to
characterize it by the observed response from the assumed standard
candle, the Crab Nebula. The different configurations of the camera
and the resultant peak response energies are summarized in
Table~\ref{configurations}.

\clearpage
\begin{deluxetable}{lcccccc}
\tablewidth{0pt}
\tablecaption{\label{configurations}Camera configurations from 1995 - 2001.}
\tablehead{\colhead{Period}&\colhead{No. of}&\colhead{Spacing}&\colhead{FOV\tablenotemark{a}}&\colhead{Light}&\colhead{Trigger}&\colhead{\epp\tablenotemark{b}}\\
                  &\colhead{Pixels}&\colhead{(\dgg)}&\colhead{(\dgg)}&\colhead{Cones}&          & \colhead{(GeV)}}
\startdata
1995/01 - 1996/12 & 109    & 0.259    & 3.0     & yes    & majority & 300            \\
1997/01 - 1997/06 & 151    & 0.259    & 3.3     & yes    & majority & 350            \\
1997/09 - 1998/12 & 331    & 0.24     & 4.8     & no     & majority & 500            \\
1999/03 - 1999/06 & 331    & 0.24     & 4.8     & yes    & pattern  & 500            \\
1999/09 - 2000/07 & 490    & 0.12\tablenotemark{c} & 3.8\tablenotemark{d} & yes    & pattern  & 430            \\
2000/10 - 2001/06 & 490    & 0.12\tablenotemark{c} & 3.8\tablenotemark{d} & yes    & pattern  & 390            \\
\enddata
\tablenotetext{a}{Field of view (FOV).}
\tablenotetext{b}{The peak response energy (\epp); this is the energy at which the
collection area folded with an E$^{-2.5}$ spectrum reaches a
maximum. Thus, it is the energy at which the camera is most efficient
at detecting \grss. These values are subject to a $\sim$ 20\%
uncertainty. Although \ep has increased somewhat over time, this does 
not mean that the camera is now poorer at detecting the lower energy
\grss. In fact, the collection area of the telescope at 300 GeV in the
1999 - 2000 observing season, was greater than the collection area at
this energy in the 1995 - 1996 observing season.}
\tablenotetext{c}{The spacing between the outer tubes is 0.24\dgg.}
\tablenotetext{d}{The outer rings of tubes were not used in this analysis, hence the
field of view here is effectively 2.6\dgg.}
\end{deluxetable}
\clearpage

The \C light images from each shower are analyzed off-line using a
moment analysis \citep{Reynolds:93}. The derived image parameters are
used to distinguish candidate \grs from the large background of
cosmic-rays. These parameters include {\it{length}}, {\it{width}},
{\it{distance}} (from the optic axis), and orientation
({\it{alpha}}). In addition, the two highest recorded signals in
individual pixels ({\it{sig1}} and {\it{sig2}}) are noted as well as
the total light in the image ({\it{size}}). Monte Carlo simulations
were used to determine the approximate limits on the parameters to be
used for the identification of candidate \gr images; these limits were
then optimized on contemporaneous observations of the Crab Nebula. The
results of an analysis are graphically presented as a histogram of the
{\it{alpha}} parameter - usually referred to as an alpha plot. In such
histograms, the {\it{alpha}} values for all the events that passed all
\gr selection criteria except for the {\it{alpha}} cut, are plotted. 
For a \gr source, an excess would be expected at low values of
{\it{alpha}}.

Two modes of observation were used: \onoff and \trackingg. In \onoff
mode, a 28 sidereal minute run is taken with the candidate \gr source
at the center of the field of view - the \on run. The \off run, a
'control' run, is also taken for 28 sidereal minutes through the same
range of azimuth and elevation, thus enabling this region of the
atmosphere to be characterised in the absence of the \gr candidate. In
the \tracking mode, the object is continuously tracked for 28 sidereal
minutes with no control run being taken.

Even in the absence of a \gr source, a certain percentage of events
recorded will pass all of the \gr selection criteria. This background
level of events, which depends on a number of factors including sky
brightness, elevation, and weather, needs to be established in order
to calculate the statistical significance of any apparent excess of
\grs detected. There are different methods to estimate this background
depending on which mode of observation was used.  

For data taken in \onoff mode, the \off scan provides an estimate of
the number of \gr like events that would be recorded from this range
of azimuth and elevation, in the absence of the candidate \gr source.
In this data acquisition mode, differences in night-sky brightness
between the \on and \off regions of the sky can introduce a bias when
the data are analyzed. In order to reduce this bias, the standard
deviations of the night-sky background in each PMT are compared for
the \on and the \off runs. Gaussian noise is then added in quadrature
to the signal from whichever region of the sky (\on or \offf) is the
darker, so as to match the standard deviations for the tube in the \on
and \off runs. This technique is known as software padding and is
described in \citet{Cawley:93}.

Unlike observations taken in the \onoff mode, scans taken in the
\tracking mode do not have independent control data. A background
estimate is essential in order to predict what number of the
background events, passing all cuts, would have been detected during
the scan in the absence of the candidate \gr source. Since most of the
\h observations were taken in the \tracking mode, two different
methods of background estimation were used. Each of these methods is
outlined below.

In the first method, which is the standard analysis method used for
data taken in the \tracking mode \citep{Catanese:98}, the alpha plot
was characterized when there was no \gr source at the center of the
field of view. This was done by analyzing 'darkfield data' which
consisted of data taken in the \off mode, and of observations of
objects found not to be sources of \grss. A large database of these
scans were analyzed, giving the shape that an alpha plot would be
expected to have when no \grs are present. 

Most \grs from an object at the center of the field of view will have
small values of the \al parameter. Hence, the \al distribution beyond
20\dg can be assumed to be independent of the \gr source, and thus
representative of the background level of \hbox{\grr-like} events in
the field of view. Using the darkfield data, a ratio was calculated to
scale the number of events between 20\dg and 65\dg to the number that
pass the \al cut. This ratio, the 'tracking ratio' ($\rho$), was then
used to scale the 20\dg to 65\dg region of the alpha plot for the
\tracking scan, to estimate the background level of events passing all
cuts. Its value and statistical uncertainty ($\Delta\rho$) are given
by:
\begin{equation}
\rho \pm \Delta\rho = \frac{N_{alpha}}{N_{control}} \pm \sqrt{\frac{N_{alpha}}{N_{control}^2} + \frac{N_{alpha}^2}{N_{control}^3}}
\end{equation}
where $N_{alpha}$ is defined as the number of events in the darkfield
data that pass all the \gr selection criteria including the \al cut,
while $N_{control}$ is the number of such events with \al between
20\dg and 65\dgg. The tracking ratio was calculated using the
darkfield data available for each observing year over the same range
of elevation angles as the source observations. In addition, the
tracking ratio was checked using the \off data on H1426+428 from
\onoff runs (where available); within statistics these tracking ratios
were consistent with the standard tracking ratio derived from the full
yearly database. To check for systematic effects associated with the
determination of the global tracking ratio, the darkfield data were
subdivided on the basis of the region of the sky which they were
from. Each of these dark-sky regions was checked individually and the
tracking ratio determined and compared to the global tracking
ratio. It was found that the global tracking ratio and these
individual tracking ratios were consistent within statistical
uncertainty.

In the alternative method for estimating the background, each \h
\tracking scan was assigned a suitable \off scan as its
background. The \off database comprised all \off scans taken during
the observing season including \h \off runs. In order to establish the
most suitable \off scan for each \h \tracking scan, each \off run was
characterized by parameters pertinent to the conditions during the
scan. These consisted of the elevation, the number of PMTs switched
off and the mean night-sky background during that scan, along with the
throughput factor recorded on the night of the scan and the date on
which the scan was taken. The throughput factor \citep{Holder:01} is a
measure of sky clarity based on the spectrum of the total amount of
light produced by background cosmic rays. The same information was
assembled for each \h \tracking scan. Then, for each \h \tracking
scan, the \off scan whose conditions most closely matched it was
deemed the most suitable background estimate, and was used as the
background for that \h \tracking run. In cases where there was not a
suitable \off run, the \h \tracking run was omitted. With each \h
\tracking scan then having an associated \off run, the data were
analyzed as if they were taken in the \onoff mode, and hence, software
padding was applied. The sky quality changes from night to night,
causing variations of $\sim 10\%$ in the raw trigger rate, for data
taken on the same source at the same elevation. Therefore, even after
selecting \off scans with run conditions which match those of the \on
scans, it is still necessary to normalize the total number of events
\on and \off. We calculate the normalizing factor from the ratio of the
number of events in the 30\dg - 90\dg control region of the \on and
\off alpha plots.

It was found that both methods of background estimation were
consistent, and indicated excesses of similar significance, thus
suggesting that the \gr excess can be reliably determined for
\tracking scans.

\section{Database}

For the purpose of data analysis, two different strategies were
employed: (a) All of the data taken in the \on and the \tracking modes
were combined and subjected to a \tracking analysis. (b) All of the
data taken in the \on mode, along with their corresponding \off data,
were analyzed together using the \pairs analysis. The \gr selection
criteria, including the {\it{alpha}} bound, were re-optimized each
year using data from the Crab Nebula. The optimum value for the
{\it{alpha}} bound was found empirically to increase from 10\dg to
15\dg when the smaller, high resolution 490 pixel camera was installed
in the summer of 1999.

\subsection{1995 - 1998 Observations}

In June-July, 1995, H1426+428 was observed for 3.5 hours in the
\tracking mode with the 109 pixel camera (spacing 0.259$^\circ$, 
3.0$^\circ$ field of view and standard trigger). In February-June,
1997, another 13.2 hours were used to observe H1426+428 in the
\tracking mode; the camera configuration was as before but the number
of pixels was increased to 151. In April, 1998, 0.87 hours of
H1426+428 data were taken; the 331 pixel camera was installed at this
time with the standard trigger.

\subsection{1999 Observations}

During the 1999 observing season, a total of 7 \onoff pairs and 51
\tracking runs were taken with the 2-fold Pattern Selection Trigger.
Three \tracking runs were excluded from further analysis based on
fluctuations in the raw rate and inferior weather conditions. This
left a total of 24.35 hours (7 \on runs and 48 \tracking runs) of data
available for analysis. These observations were initiated as part of
the general BL Lac survey (D. Horan et al., in preparation) with extra
data being taken in 1999 because of an initial indication of a signal
from \hh. A tracking ratio of 0.232 $\pm$ 0.005 was determined for \al
$<$ 10\dgg.

\subsection{2000 Observations}

During the 2000 observing season, 33 \onoff pairs and 35 \tracking
runs were taken on H1426+428. Eleven \on source runs were excluded
from further analysis based on fluctuations in the raw rate and
inferior weather conditions, leaving a total of 26.46 hours of data
available for \tracking analysis (57 \trackingg/\onn) and 13.86 hours
of data in the \onoff mode (30 \onoff pairs). A tracking ratio of
0.312 $\pm$ 0.002 for {\it{alpha}} $<$ 15\dg was
determined. H1426+428 \off data were analyzed to ensure that this
tracking ratio was consistent with the tracking ratio derived for
these data.

\subsection{2001 Observations}

A total of 42.24 hours were spent on the source during 2001; this
comprised 39 \onoff pairs and 59 \tracking runs. After excluding
data taken at low elevation and in unsatisfactory weather
conditions, 38.10 hours of data remained (34 \onoff pairs and 53
\tracking runs). The tracking ratio for this observing season, again 
for {\it{alpha}} $<$ 15\dgg, was calculated to be 0.323 $\pm$
0.002. As with the 2000 data, checks were performed to ensure that
this ratio was appropriate for the H1426+428 data. A summary of the
data taken and analyzed in the 2000 and 2001 observing seasons is give
in Table~\ref{data}.

\clearpage

\begin{table}
\begin{center}
\caption{Summary of the number of data scans taken and analyzed in 2000 and 2001.\label{data}}
\begin{tabular}{ccccccc}
\tableline\tableline
Year & \multicolumn{2}{c}{Data Recorded}   & \multicolumn{3}{c}{Data Analyzed}                      \\
     & \tracking                & \on      & Standard       & Alternative         & \pairs          \\
\tableline
2000 & 35 (27)                  & 33 (30)  & 57             & 57                  & 30     \\
2001 & 59 (53)                  & 39 (34)  & 87             & 71                  & 34     \\
\tableline
\end{tabular}
\tablecomments{The mode that the data were recorded in are given in 
the first two columns. The numbers in parentheses here refer to the
number of usable data scans recorded. The final three columns give the
number of scans analyzed in each of the three data analysis modes:
standard \tracking analysis, alternative \tracking analysis or
\pairss.}
\end{center}
\end{table}

\section{Results}

\subsection{1995 - 1998 Observations}
The observations of H1426+428 between 1995 and 1998 were analyzed as
part of a general BL Lac survey. The results are summarized in
Table~\ref{95-99}. Although not statistically significant, \h was one
of only two objects, out of the 24 objects observed in this survey, to
show consistently positive results.

\clearpage
\begin{deluxetable}{cccccc}
\tablewidth{0pt}
\tablecaption{Summary of the Observations and Results for H1426+428 between 1995 and 1999.\label{95-99}}
\tablehead{ &\colhead{Exp.}&\colhead{Total}&\colhead{Max. $\sigma$}&\colhead{Max. $\sigma$}&\colhead{Flux\tablenotemark{c}}\\
\colhead{Period}&\colhead{(hrs)}&\colhead{$\sigma$}&\colhead{Month\tablenotemark{a}}&\colhead{Night\tablenotemark{b}}&\colhead{(cm$^{-2}$ s$^{-1}$)}}
\startdata
1995/06 - 1995/07 &  3.48 & 2.1      & 2.1                   & 2.1                   & $<$2.2 x 10$^{-11}$  \\
1997/02 - 1997/06 & 13.16 & 1.7      & 2.2                   & 1.6                   & $<$1.6 x 10$^{-11}$  \\
1998/04           &  0.87 & 1.7      & 1.7                   & 2.0                   & $<$6.7 x 10$^{-11}$  \\
1999/03 - 1999/06 & 24.35 & 0.9      & 1.6                   & 2.1                   & $<$6.7 x 10$^{-11}$  \\
\enddata
\tablenotetext{a}{The maximum statistical significance of the signal 
recorded from \h when the data are grouped by the month during which
they were recorded.}
\tablenotetext{b}{The maximum statistical significance of the signal 
recorded from \h when the data are grouped by the night on which they
were recorded.}
\tablenotetext{c}{The integral fluxes are quoted at the peak response energy
for the observation period, as given in Table~\ref{configurations}.}
\end{deluxetable}
\clearpage

\subsection{1999 Observations}

A total of 24.35 hours of data were taken on \h during 1999. The net
excess of +0.9$\sigma$ from the combined 1999 observations was not
statistically significant.

\subsection{2000 Observations}

\begin{figure}[h!]
\epsscale{1}
\plottwo{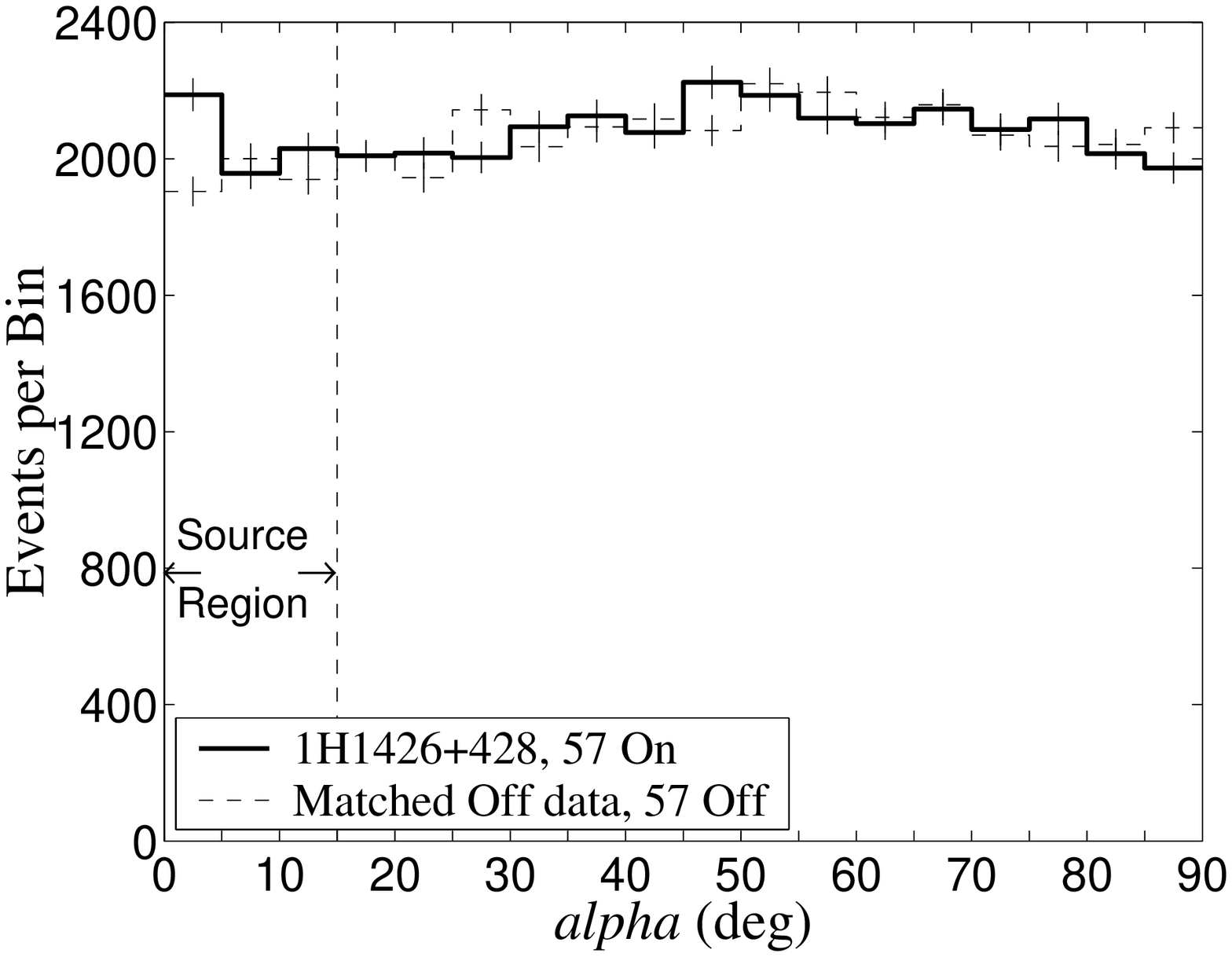}{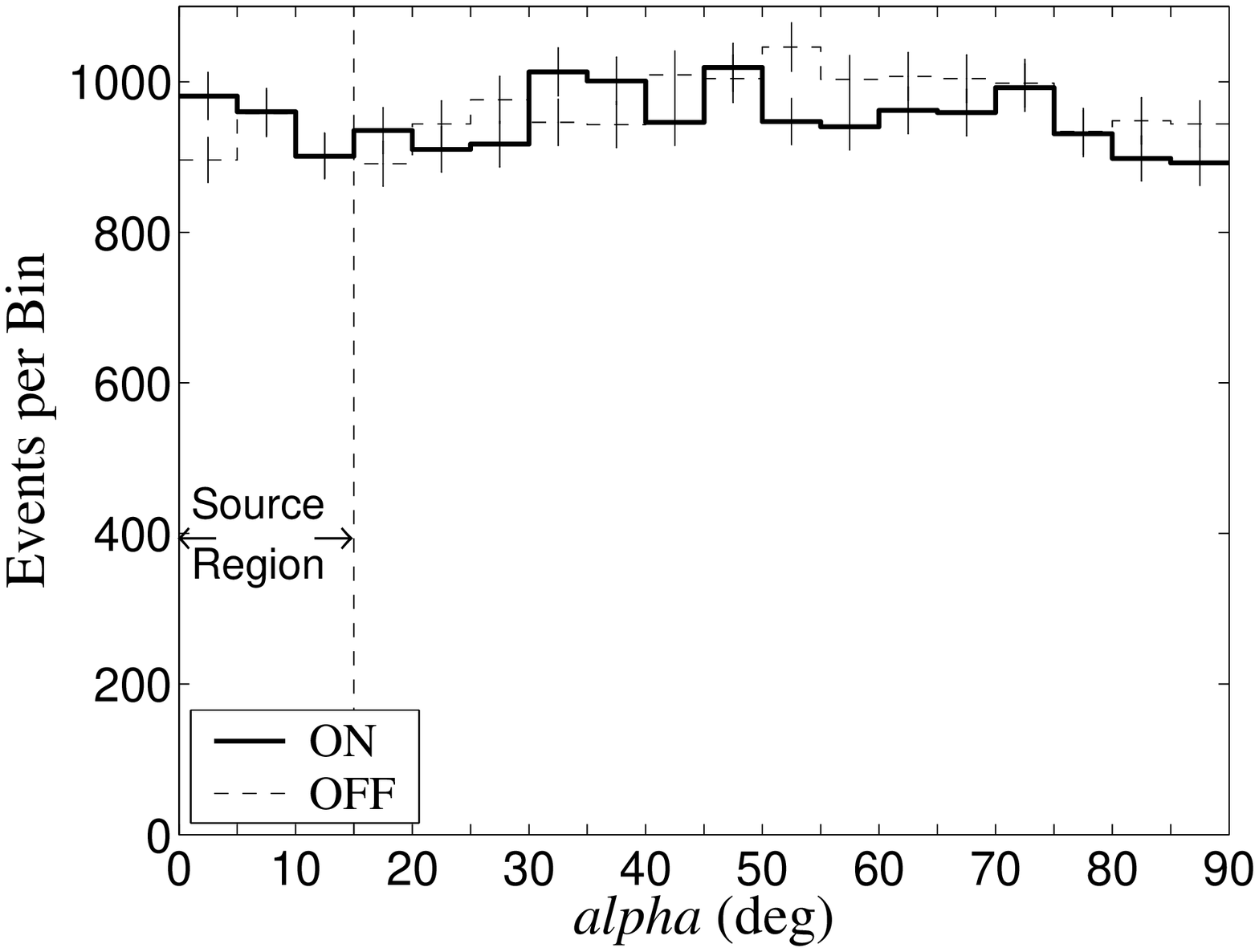}
\caption {\label{alpha-00}Left: Alpha plot for 26.49 hours
of the data taken on \h during 2000. The total significance for these
57 scans is +3.1$\sigma$ when analyzed using the alternative method
for estimating the background. The matched \off data used to estimate
the background level are also shown (dashed line). These are
normalized to the \h data between {\it{alpha}} values of 30\dg and
90\dgg.  Right: Alpha plot for the 30 \onoff pairs taken in 2000. The
net significance for these data is +1.2$\sigma$.}
\end{figure}

A \tracking analysis of all the on-source data taken on \h during
2000, using the standard tracking ratio calculated for that observing
season, gave a +4.2$\sigma$ excess corresponding to a \gr rate of 0.24
$\pm$ 0.06 \gsp min$^{-1}$. When these data were analyzed
independently using the alternative method of background estimation,
the excess was at the +3.1$\sigma$ level with a \gr rate of 0.21 $\pm$
0.07 \gsp min$^{-1}$. The alpha plot for these data is shown in
Figure~\ref{alpha-00} (left panel) along with the alpha plot from the
matched \off data. The background shown on the figure (dashed line)
consists of the \off data which were chosen to match the
characteristics of the \h \tracking data. These data are scaled to
match the \h data in the 30\dg - 90\dg region of the alpha plot.  The
net significance for the 30 \onoff pairs taken in 2000 was
+1.2$\sigma$, with a \gr rate of 0.10 $\pm$ 0.09 \gsp min$^{-1}$. The
alpha plot for these data is shown in Figure~\ref{alpha-00} (right
panel). On the night of May 30, 2000 (MJD 51694), the first
\tracking scan on \h gave a significance of +3.0$\sigma$, using the
standard tracking ratio method of background estimation. This prompted
observers to take four more \tracking scans which led to the detection
of a signal at the +3.7$\sigma$ for this night. The \gr rate for these
5 scans (2.15 hours) was 0.67 $\pm$ 0.18 \gsp min$^{-1}$. 

\clearpage
\clearpage

\subsection{2001 Observations}

\begin{figure}[h!]
\epsscale{1}
\plottwo{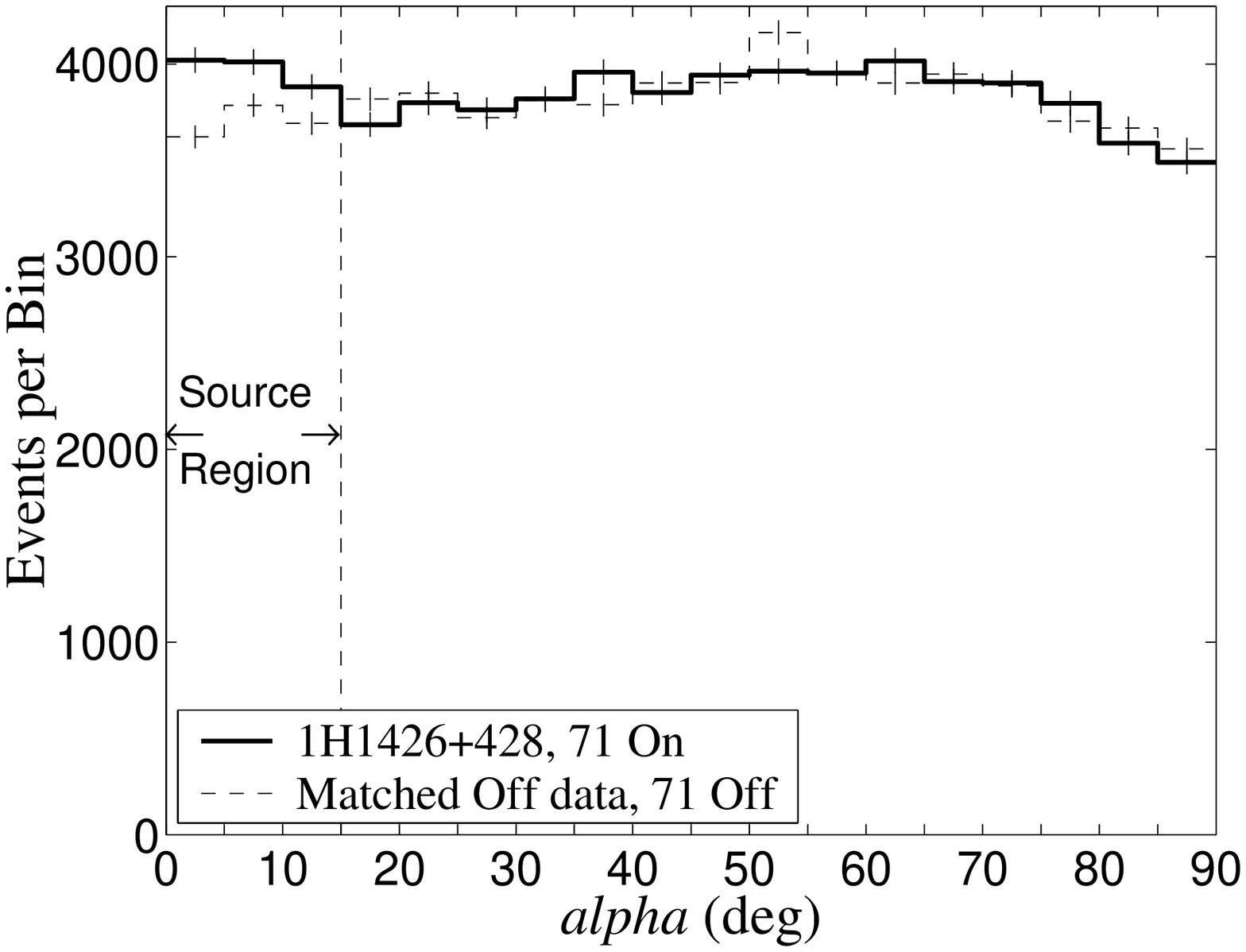}{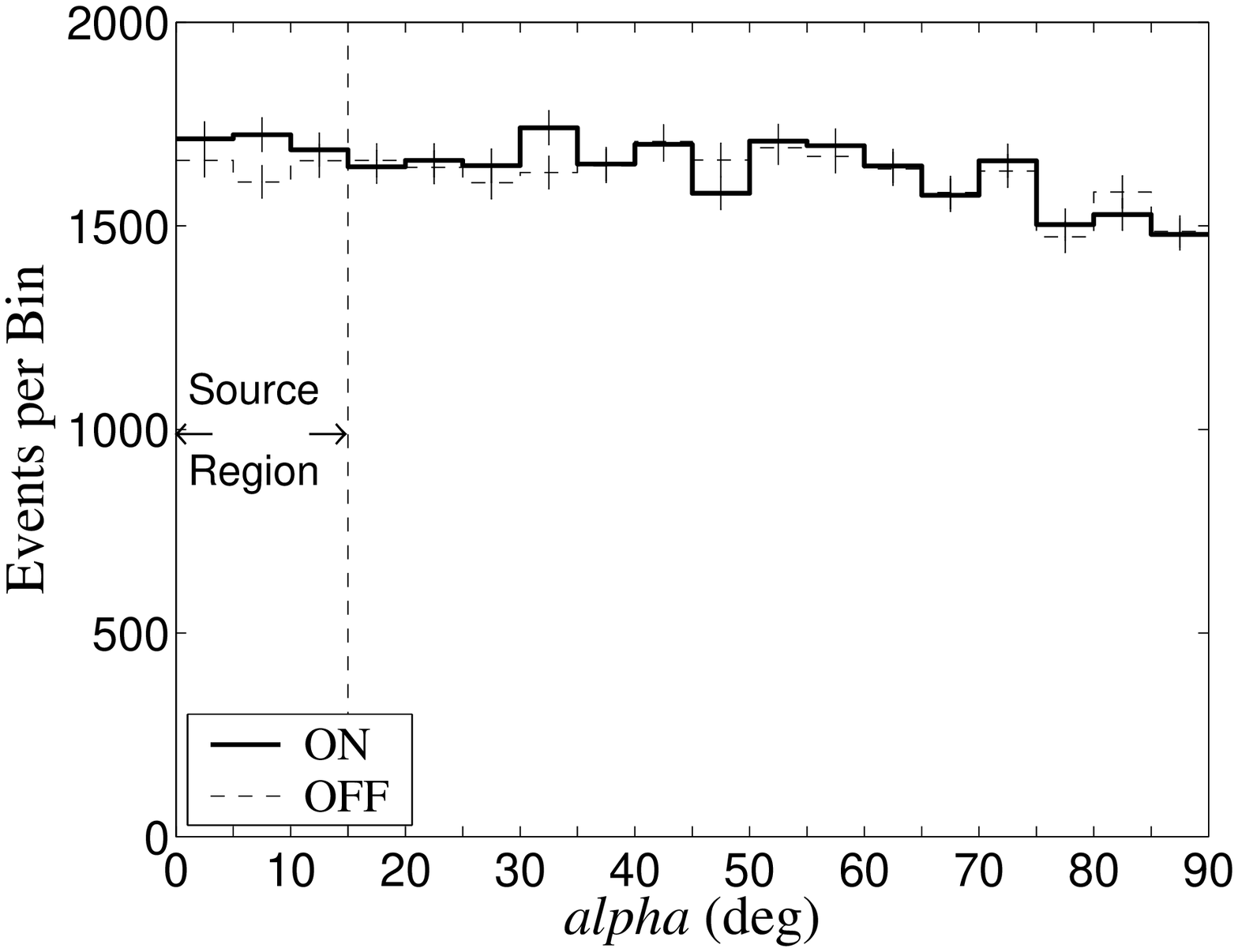}
\caption {\label{alpha-01} Left: Alpha plot for 31.13 hours
of the data taken on \h during 2001. The total significance for these
71 scans, when analyzed using the alternative method of background
estimation, is +5.5$\sigma$. The matched \off data used to estimate
the background level are also shown (dashed line). These are
normalized to the \h data between {\it{alpha}} values of 30\dg and
90\dgg. Right: Alpha plot for the 15.53 hours of data taken in the
\onoff mode on \h during 2001. The total significance for these 34
\onoff pairs is +2.0$\sigma$.}
\end{figure}

A \tracking analysis of all of the on-source data taken on \h during
2001, using the tracking ratio calculated for this observing season,
gave an excess of +5.4$\sigma$ corresponding to a \gr rate of 0.36
$\pm$ 0.07 \gsp min$^{-1}$. An independent analysis of 31.13 hours of
these data, using the alternative method of background estimation,
resulted in a +5.5$\sigma$ excess with a \gr rate of 0.44 $\pm$ 0.08
\gsp min$^{-1}$. The alpha plot for these data is shown in
Figure~\ref{alpha-01} (left panel) along with the alpha plot from the
matched \off data. As before, these data are scaled to the \h data
between {\it{alpha}} values of 30\dg and 90\dgg.

\clearpage
\clearpage

The significance from the 34 \onoff pairs taken during the 2001
observing season is +2.0$\sigma$, with a \gr rate of 0.22 $\pm$ 0.11
\gsp min$^{-1}$. The alpha plot for these data is shown in
Figure~\ref{alpha-01} (right panel). A summary of the results of the
observations made on \h during 2000 and 2001, using the alternative
method of background estimation, is presented in Table~\ref{sum-res}.

\clearpage
\begin{deluxetable}{rcccc}
\tablewidth{0pt}
\tablecaption{\label{sum-res}Summary of the results of H1426+428 observations during 2000 and 2001.}
\tablehead{\colhead{Period}&\colhead{Exp. (hrs)} &\colhead{Total $\sigma$}&\colhead{Max. $\sigma$ Month\tablenotemark{a}}}
\startdata
2000/02 - 2000/06         &26.37 & 3.1       & 3.4                   \\
2001/01 - 2001/06         &31.12 & 5.5       & 5.0                   \\
\enddata
\tablenotetext{a}{The maximum statistical significance of the signal 
recorded from \h when the data are grouped by the month during which
they were recorded.}
\end{deluxetable}
\clearpage

When both the differences in exposure time and methods of background
estimation are taken into account, the results from the \onoff pairs
and the \tracking data taken on \h during each observing season are
consistent, at the 2$\sigma$ level, with each other.

\subsection{Comparison of the Gamma-Ray and X-Ray Flux from
H1426+428}

The \gr rates for the 2000 and 2001 \h data were compared with the
x-ray flux from the All Sky Monitor (ASM) instrument on board the
Rossi X-ray Timing Explorer (RXTE; \citealp{Levine:96}). A
correlation was sought between the nightly \gr rates and the
``one-day average'' x-ray data points. Only nights on which there
were both x-ray and \gr data were considered when performing the
correlation. This left 37 nights for analysis during 2000, and 39
nights during 2001.

\begin{figure}[h!]
\epsscale{0.8}
\plotone{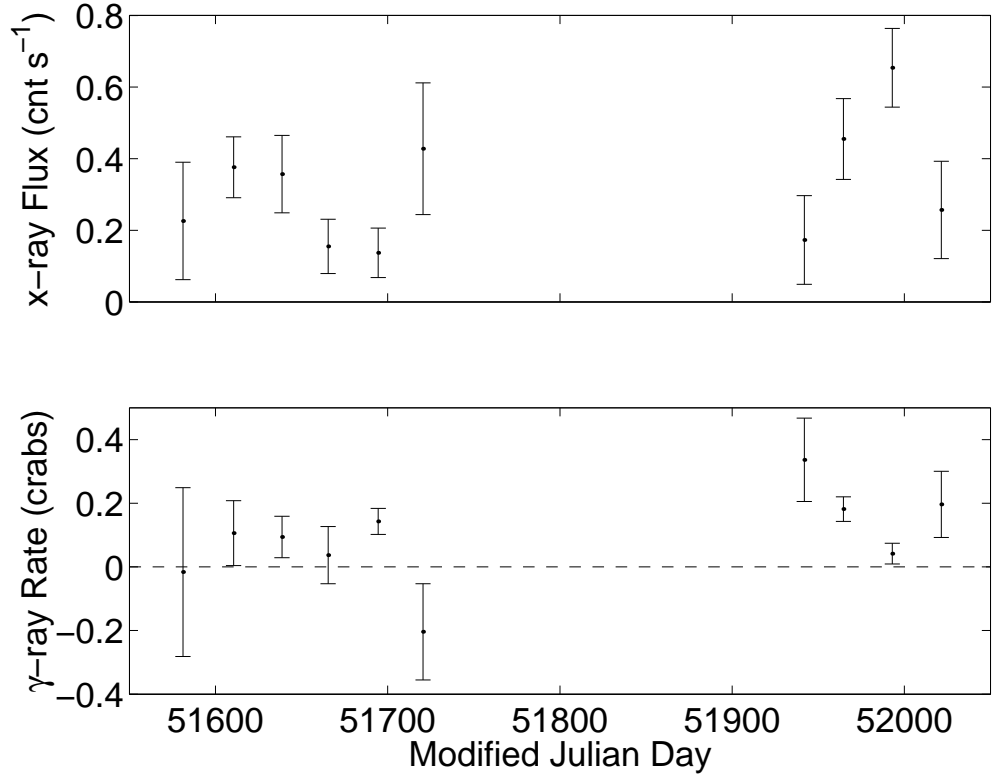}
\caption {\label{xray} Top: The mean x-ray rate for \h for each 
month during 2000 and 2001 from the ASM on RXTE. The data are only
plotted for months during which \gr data were also taken. Bottom: The
mean \gr rate from \h for each month calculated using the alternative
method for background estimation. This rate is plotted in units of the
\gr rate from the Crab for that year. The detected Crab rate was 2.45
\gsp min$^{-1}$ during 2000 and 3.36 \gsp min$^{-1}$ during 2001.}
\end{figure}

Since \gr observations can only be taken on moonless nights, there are
a few ($\sim$ 6) nights around the time of full moon each month, when
no observations can be made. The periods during which the \gr data are
taken are referred to as 'darkruns'. Correlations between the x-ray
and \gr rates were sought both over the entire observing season, and
during darkruns with more than two nights on which both x-ray and \gr
data were taken. No evidence for significant nightly correlation was
found for either observing season. The x-ray and \gr rate curves are
shown in Figure~\ref{xray}. The rates plotted here are the average
monthly rates for the 2000 and 2001 data. The \gr rates shown were
calculated using the alternative method of background
estimation. There is some evidence for an anti-correlation between
these average monthly rates, especially in the 2001 data.

\clearpage
\clearpage

\section{Spectral Characteristics of H1426+428}
\label{spectrum}

The TeV flux from H1426+428 is weak and near the threshold of
sensitivity of the 10m \gr telescope.  The photon flux is so small
that it is impractical to apply standard spectral analysis techniques,
for example, \citet{Mohanty:98}, to these data. In fact, only the
differential flux from the source for the energy at which the
telescope is most efficient at detecting \grss, can be evaluated with
reasonable accuracy.  Such a flux estimate accounts mostly for the
photons with energies corresponding to the peak in the differential
detection rate of the telescope.  This energy, E$_p$, depends on the
spectral index of the source in question. For observations of \h with
the Whipple telescope in 2001, $E_{p}$ was found to be between 280 -
360 GeV. The rate of change of the number of excess events, presumably
photons from the source, with the total amount of light in the \C
image, the {\it size}, is directly related to the spectral index of
the source. This relationship is established using Monte Carlo
simulations; the spectral index of the source in the vicinity of
$E_{p}$ is then evaluated. This in turn allows estimates of $E_p$ for
this spectral index, and the differential and integral fluxes to be
improved by reducing the uncertainty due to the unknown spectral
index.

\begin{figure}[h!]
\epsscale{0.8}
\plotone{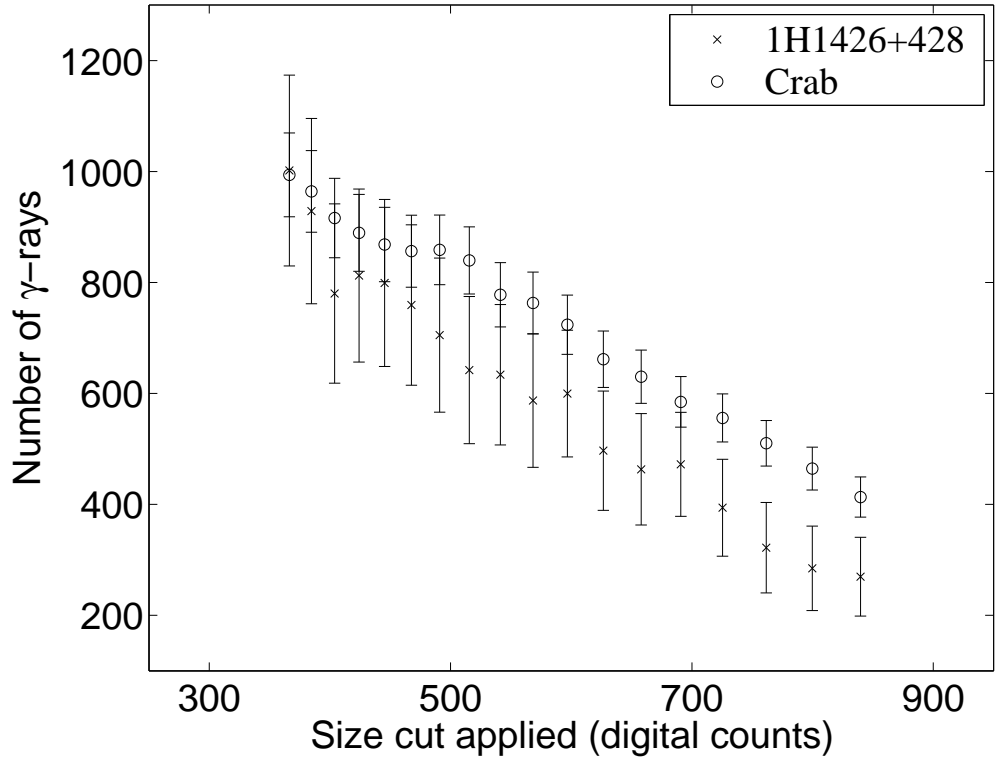}
\caption{\label{cr-1h}Integral excess events observed by the Whipple telescope
from the directions of H1426+428 (crosses) and the Crab Nebula (open
circles) during 2001 as a function of integrated \C light in
the shower image. Exposure on the Crab Nebula was adjusted to match
the total excess of H1426+428 at the lowest {\it{size}} cut applied,
366 digital counts. One photoelectron corresponds to $\sim$ 3.6
digital counts.}
\end{figure}

Figure~\ref{cr-1h} shows the excess events detected from the direction
of H1426+428 as a function of integrated \C light in the shower
image. For comparison, this plot also shows the excess events detected
from the Crab Nebula after a 4.1 hour exposure, which was chosen to
produce a similar maximal excess of $\sim$ 10$^3$ events as was found
during the 32.5 hours of observation of H1426+428 during 2001. The
lowest {\it size} cut of 366 digital counts is the result of signal to
noise optimization of the \h data, producing an overall significance
close to +6$\sigma$.  The highest {\it size} cut, 840 digital counts,
is limited by the rapidly declining significance to the level of
+3.5$\sigma$.  To a first approximation, the dependence of the excess
events on the logarithm of the {\it{size}} cut applied, is a linear
function whose slope is monotonically related to the spectral index of
the source.  This plot indicates that the H1426+428 spectrum is
softer than the well-studied spectrum of the Crab Nebula.

Monte Carlo simulations were used to relate the observed parameters of
linear fits to the data in Figure~\ref{cr-1h}, to the spectral
characteristics of H1426+428, and the Crab Nebula.  The statistical
correlation of the errors shown in this integral plot of excess events
has been taken into account to derive optimal estimates of the
spectral indices, differential and integral fluxes, and their
errors. A summary of the results is given in Table~\ref{spectral}. For
the Crab Nebula a spectral index of 2.75 with an uncertainty of $6$\%
(1$\sigma$) was derived. The peak of the differential detection rate
of the Whipple telescope from a source with such a spectral index has
been found to be around 360 GeV (2001 observing season). Based on the
small Crab Nebula data-set, the spectral index as well as the
differential and integral fluxes have been derived, and are found to
be consistent, at the 2$\sigma$ level, with our previous observations
\citep{Hillas:98}. Analysis of the H1426+428 data-set indicates a
substantially steeper spectral index of 3.55 with a 13\% relative
error. For this spectral index and the 366 digital counts {\it{size}}
cut applied in the data analysis, the differential photon detection
rate peaks at $\sim$ 280 GeV.  Table~\ref{spectral} shows estimates of
the fluxes at 390 GeV assuming the same spectral index. This energy
was chosen by increasing the {\it size} cut until the signal to noise
ratio decreased to $\sim$ 4.5 $\sigma$; below this value the signal
was not considered strong enough for reliable calculation of the \h
flux.

\clearpage
\begin{deluxetable}{cccc}
\tablewidth{0pt}
\tablecaption{\label{spectral}Results of the spectral analysis of H1426+428 and Crab Nebula data-sets.}
\tablehead{
\colhead{Parameter}                          & \colhead{Crab}      & \colhead{H1426+428}& \colhead{H1426+428}}
\startdata
$\chi^2$                                     & 20.60               & 22.72               & 22.72            \\ 
d.f.                                         & 16                  & 16                  & 16               \\ 
T (s)                                        & 14.7 x 10$^{3} $    & 117.3 x 10$^{3} $   & 117.3 x 10$^{3}$ \\ 
\hline
$\alpha$                                     & 2.75                & 3.55                & 3.55             \\
$E_p$ (GeV)                                  & 360                 & 280                 & 390              \\
$F_p$ (cm$^{-2}$ s$^{-1}$)                   & 1.10 x 10$^{-10}$   & 2.04 x 10$^{-11}$   & 8.76 x 10$^{-12}$\\
$dF_p$ (cm$^{-2}$ s$^{-1}$ TeV$^{-1}$)       & 5.34 x 10$^{-10}$   & 1.86 x 10$^{-10}$   & 5.73 x 10$^{-11}$\\ 
$\nu F_{\nu}$ (erg cm$^{-2}$ s$^{-1}$)       & 1.11 x 10$^{-10}$   & 2.33 x 10$^{-11}$   & 1.39 x 10$^{-11}$\\
$\frac{\Delta \alpha}{\alpha}$               & 0.06                & 0.13                & 0.13             \\ 
$\frac{\Delta F_p}{F_p}$                     & 0.07                & 0.17                & 0.23             \\ 
$\frac{\Delta dF_p}{dF_p}$                   & 0.11                & 0.25                & 0.17             \\ 
\enddata
\tablecomments{$\alpha$ is the spectral index of the differential flux.
$E_p$ denotes the energy at which the differential detection rate of
photons from a given source, with given observation and data analysis
conditions, peaks.  $F_p$ and $dF_p$ are estimates of the integral and
differential fluxes at $E_p$.}
\end{deluxetable}

\clearpage

\clearpage

\begin{figure}[h!]
\epsscale{0.8}
\plotone{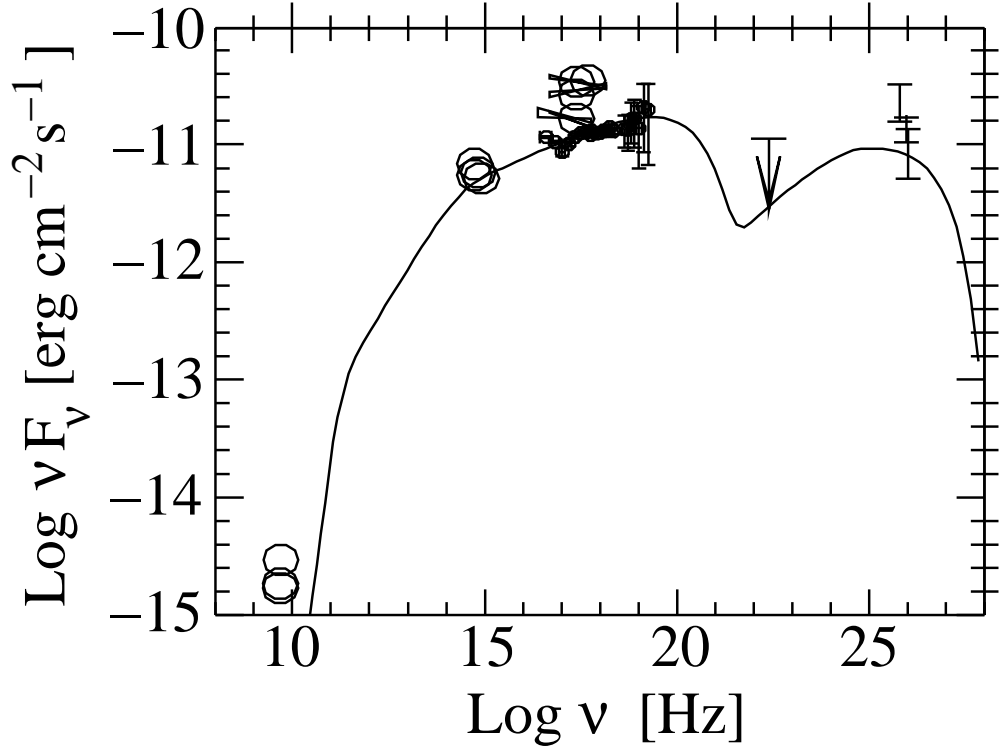}
\caption {\label{fig-SED}SED of \h from \citet{Costamante:01}, along 
with the observational results obtained with the Whipple
telescope. The 280 GeV (log $\nu$ = 25.83) and 390 GeV (log $\nu$ =
25.97) data points, derived from 2001 data-set, suggest a differential
spectral index of the energy flux equal to -3.55. The 430 GeV (log
$\nu$ = 25.99) point is a detection by the Whipple telescope during
2000.  This result is subject to a large uncertainty associated with
the weak detection and unknown spectral properties of the source. The
EGRET upper limit above 100 MeV of 0.7 x 10$^{-7}$ cm$^{-2}$ s$^{-1}$
is also shown as are the fluxes from various x-ray, optical, and radio
experiments, all taken from \citet{Costamante:01}.}
\end{figure}

The observed energy fluxes of H1426+428 at $280$ and $390$ GeV are
shown in Figure~\ref{fig-SED} together with the predicted SED for this
source suggested by \citet{Costamante:01}. This figure also shows the
tentative detection (+3.1 $\sigma$) made by the Whipple 10m telescope
during 2000. Operating at a substantially higher peak response energy
at that time, the Whipple telescope observed this source at \ep $\sim$
430 GeV.  Based on these data an estimate of the energy flux of 0.86
$\pm$ 0.33 $\times$ 10$^{-11}$ erg cm$^{-2}$ s$^{-1}$ has been
derived. Due to the very weak signal however, we have not attempted to
find the spectral index of H1426+428 for this data-set.  As a result,
the indicated error is only statistical and it does not include errors
associated with the unknown spectral properties of the source. The
$430$ GeV peak energy itself is subject to a similar uncertainty.

\clearpage
\clearpage

\section{Discussion}

Observations of H1426+428 with the Whipple 10m telescope since 1995
have returned consistently positive excesses although the statistical
significance has not always been high. H1426+428 was on the very
short list of predicted TeV emitters published by the \bpsp
collaboration \citep{Costamante:00a} and was also singled out as the
most probable TeV emitter amongst these due to the high frequency of
its synchrotron peak. It was observed as part of a BL Lac survey at
the Whipple Observatory between 1995 and 1998, and was singled out for
more extensive observations between 1999 and 2001. The strongest
evidence for VHE emission was recorded during the 2000 and 2001
observing seasons, with signals at the 3.1$\sigma$ and 5.5$\sigma$
levels, respectively, being detected. Although it is difficult to
combine results from different configurations of the camera, the
combined significance of the results from 2000 and 2001 is clearly
equivalent to a greater than 5$\sigma$ detection, the usual level
required to establish the existence of a new TeV source.

There have been two occasions (in 1999 and in 2000) when there was
marginal evidence for a transient signal from H1426+428. In 1999,
after 2.3 hours of observations (over two nights) a signal at the
+3.1$\sigma$ level was recorded while in 2000, a signal at the
+3.7$\sigma$ level was recorded after 2.2 hours of observations (on a
single night).

After 18.1 hours of observations on H1426+428, the CAT collaboration
reported a 3$\sigma$ upper limit of 3.54 x 10$^{-11}$ cm$^{-2}$
s$^{-1}$ at an energy threshold of 250 GeV \citep{Piron:00}. These
observations were carried out at the beginning of February 1999 in
response to an announcement from the \bpsp and the RXTE/ASM teams,
which revealed H1426+428 to be transitioning to a high state.

If independently verified, the observations reported here represent
an important addition to the catalog of TeV-emitting BL Lacs 
(Table~\ref{catalog}). That the source was predicted to be a TeV
\gr source based on its x-ray spectrum is important in that
it signifies the maturity of the observational techniques and the
theoretical understanding of BL Lacs. It reinforces the symbiosis
between observations at x-ray wavelengths, particularly in hard
x-rays, and those at TeV energies, particularly those with good
sensitivity below energies of 1 TeV. The existence of a population of
sources whose most prominent emission is at energies of 10-100 keV and
300-1000 GeV points to a fruitful overlap between the next generation
of ground-based atmospheric \C telescopes \citep{Weekes:01b}
and the future hard x-ray experiment, EXIST \citep{Grindlay:00}. It is
remarkable also in that this source, like Mrk 501 and 1E2344+514, is
not included in the 3rd EGRET Catalog, supporting the interpretation
of \citet{Ghisellini:99} that in blazars sufficiently powerful to be
detected readily by EGRET, the spectrum will terminate at lower
energies and not extend to the TeV domain.

\clearpage
\begin{deluxetable}{lccccc}
\tablewidth{0pt}
\tablecaption{\label{catalog}Extra-galactic TeV Sources \citep{Catanese:99}}
\tablehead{\colhead{Source}& \colhead{Type} & \colhead{$z$} & \colhead{Discovery} & \colhead{Group} & \colhead{EGRET}}
\startdata
Markarian 421  & HBL  & 0.031 & 1992      & Whipple                                & yes         \\
               &      &       &           & \citep{Punch:92}                       &             \\
Markarian 501  & HBL  & 0.034 & 1995      & Whipple                                & yes         \\
               &      &       &           & \citep{Quinn:96}                       &             \\
1ES 2344+514   & HBL  & 0.044 & 1997      & Whipple                                & no          \\
               &      &       &           & \citep{Catanese:98}                    &             \\
1ES 1959+650   & HBL  & 0.048 & 1999      & Telescope Array                        & no          \\
               &      &       &           & \citep{Nishiyama:00}                   &             \\
PKS 2155-304   & HBL  & 0.116 & 1999      & Durham                                 & yes         \\
               &      &       &           & \citep{Chadwick:99}                    &             \\
H1426+428     & HBL  & 0.129 & 2001      & Whipple                                & no          \\
               &      &       &           & (Horan et al. 2000, 2001a, 2001b)      &             \\
3C66A          & LBL  & 0.444 & 1998      & Crimea                                 & yes         \\
               &      &       &           & \citep{Neshpor:98}                     &             \\
BL Lacertae    & LBL  & 0.069 & 2001      & Crimea                                 & yes         \\
               &      &       &           & \citep{Neshpor:01}                     &             \\
\enddata
\end{deluxetable}
\clearpage

Although the properties of H1426+428 reported here are scanty in
comparison with the other, better studied TeV BL Lacs, they agree in
principle with the characteristics now well established for Mrk 421
and Mrk 501 (and to a lesser extent, 1ES 2344+514). Although the flux
from \h is low, there is some evidence for time variability. There is
also an indication of a hard \gr spectrum. There is no evidence for a
correlation of x-ray and TeV \gr fluxes (unlike Mrk 501) but the
sensitivity of observations to date is limited and even in the
stronger sources this correlation has been shown to be complicated.

\citet{Ghisellini:99} has proposed that there is a continuous sequence
of BL Lacs with the peak of the synchrotron spectrum increasing as the
luminosity decreases. H1426+428 is another example of an "extreme" BL
Lac characterized by a synchrotron spectrum that may peak near 100 keV
and a relatively weak luminosity. These objects are the best
candidates for TeV emission based on Compton-synchrotron models of BL
Lac jets. However, the observed variations in sources such as Mrk 501
cannot be accounted for by simple one-zone homogeneous self-Compton
models and may require another source of optical target photons. It
will require the detection of a greater population of sources such as
H1426+428 to test these models fully. It has been noted
\citep{Ghisellini:98} that these TeV-emitting BL Lacs are also
characterized by a relatively strong radio luminosity and that the
radio luminosity may be a measure of the density of additional seed
photons at infrared-optical wavelengths needed to explain the TeV
emission.

This is the most distant of the TeV-detected BL Lacs classified as
HBL, and hence it has promising implications for the detection of more
BL Lacs at $z$ $>$ 0.1. Stronger detections of such sources, which
allow an accurate measure of the TeV energy spectrum, may place
significant limits on the density of the intergalactic background
light. The steep spectrum derived here is consistent with many models
of intergalactic absorption but could also be intrinsic to the source.

Interpretation of the theoretical prediction of the intrinsic spectrum
of H1426+428 and the observational results reported in this paper are
complicated by the possible strong attenuation of the high energy
photons by the diffuse intergalactic infrared background.  The
appearance of such a cutoff in the spectra of TeV extragalactic
sources was suggested in one of the pioneering works on TeV \gr
absorption, by \citet{Stecker:92}, for the quasar 3C 279 ($z$ =
0.54). A later paper (V. V. Vassiliev et al., in preparation) will
discuss this subject in detail in relation to \hh; here we note only
that at 280 GeV the attenuation optical depth can be anywhere between
$z$ of 0.1 and 1.0 depending on the extreme upper and lower limits
known for the density of the extragalactic background light. Due to
this potentially large absorption effect it is possible, for example,
that the intrinsic spectral index of this source is softened by $\sim$
1.75 to produce the observed value of 3.55, if the intrinsic
properties of H1426+428 are similar to those of Mrk 501
\citep{Vassiliev:01}. It remains certain however, that being the most
distant HBL detected at sub-TeV energies, the observed spectral
properties of \h and the understanding of its intrinsic
characteristics from multiwavelength observations will provide the
most constraining data for extragalactic background light studies.

\section{Acknowledgments}
The authors would like to thank Kevin Harris, Joe Melnick, Emmet
Roach, and all the staff at the Whipple Observatory for their
support. We also thank the referee for many constructive and useful
suggestions. This research was supported in part by the
U. S. Department of Energy, PPARC, and Enterprise Ireland.

\appendix
\section{Data Log for 2000 and 2001 Observations}

\begin{table}
\begin{center}
\caption{\label{datalog}Data log for the 2000 and 2001 \h Observations}
\begin{tabular}{ccccc}
\tableline\tableline
\multicolumn{5}{c}{MJD Start Time [Mode]} \\
\tableline
51577.48031 [p] & 51672.28898 [t] & 51941.43213 [t] & 51970.44161 [t] & 51994.37299 [p] \\
51577.52101 [t] & 51687.17778 [t] & 51941.52991 [t] & 51970.49679 [t] & 51994.41703 [p] \\
51578.49144 [p] & 51687.19735 [t] & 51942.43760 [t] & 51970.51712 [t] & 51995.49070 [p] \\
51579.47485 [p] & 51687.21689 [t] & 51944.49363 [t] & 51971.43684 [t] & 51996.43112 [t] \\
51579.51770 [t] & 51687.23145 [t] & 51957.37136 [p] & 51971.45942 [t] & 51996.47550 [t] \\
51581.48048 [p] & 51688.17238 [p] & 51957.39284 [t] & 51972.44811 [t] & 51997.35435 [p] \\
51585.50929 [t] & 51689.17450 [p] & 51959.39676 [p] & 51972.47175 [t] & 51997.44241 [t] \\
51586.50519 [t] & 51690.17385 [p] & 51959.42716 [t] & 51986.36833 [p] & 51997.46204 [t] \\
51604.43009 [t] & 51692.20786 [t] & 51959.44517 [t] & 51988.33540 [p] & 51998.40471 [t] \\
51605.40456 [p] & 51692.24942 [p] & 51960.37838 [p] & 51988.35239 [t] & 51998.44546 [p] \\
51607.43234 [p] & 51693.16773 [p] & 51960.40149 [t] & 51989.38498 [t] & 51999.41745 [p] \\
51617.42927 [p] & 51694.17954 [t] & 51960.44430 [p] & 51989.43090 [p] & 51999.45334 [p] \\
51632.35804 [t] & 51694.19409 [t] & 51960.46341 [t] & 51990.39131 [t] & 52000.40087 [t] \\
51632.37754 [t] & 51694.22455 [p] & 51960.48046 [t] & 51990.43151 [p] & 52000.44380 [p] \\
51635.31200 [t] & 51694.26542 [p] & 51960.50067 [t] & 51990.46683 [t] & 52020.23130 [p] \\
51635.33273 [t] & 51694.30629 [p] & 51961.37826 [p] & 51991.40168 [p] & 52022.57165 [p] \\
51636.35315 [p] & 51695.17682 [p] & 51961.39750 [t] & 51991.41978 [t] & 52046.22484 [p] \\
51637.30195 [p] & 51695.21768 [p] & 51961.43929 [p] & 51991.43755 [t] & 52047.31743 [t] \\
51638.35877 [p] & 51695.25854 [p] & 51961.47573 [t] & 51991.46142 [t] & 52049.27725 [t] \\
51641.30764 [p] & 51696.26489 [t] & 51961.49286 [t] & 51991.47977 [t] & 52050.19936 [p] \\
51642.36119 [t] & 51698.17804 [t] & 51962.39036 [p] & 51992.38284 [p] & 52050.23721 [p] \\
51642.37579 [t] & 51699.17421 [p] & 51962.40577 [t] & 51992.43252 [p] & 52051.19595 [p] \\
51644.36277 [t] & 51700.23519 [p] & 51962.43053 [p] & 51992.47643 [t] & 52051.26034 [p] \\
51645.36182 [p] & 51701.25211 [p] & 51963.38181 [t] & 52275.18178 [p] & 52053.23280 [p] \\
51659.25434 [p] & 51702.28313 [t] & 51963.42351 [t] & 52275.18443 [t] & 52055.26242 [p] \\
51659.29540 [t] & 51702.30271 [t] & 51963.46618 [t] & 52275.18320 [t] & 52056.30267 [p] \\
51667.28374 [p] & 51720.18600 [t] & 51963.48589 [t] & 52275.18199 [t] & 52071.24343 [t] \\
51669.26859 [p] & 51721.18547 [p] & 51965.45873 [t] & 52275.18201 [t] & 52072.21843 [t] \\
51672.26933 [t] & 51940.53791 [t] & 51967.46276 [t] & 52275.18178 [t] &                 \\
\tableline
\end{tabular}
\tablecomments{The number in parentheses after the MJD gives the mode 
in which the data were taken: p - \pairss, t - \trackingg.}
\end{center}
\end{table}

\end{document}